\documentclass{aa}  
\usepackage[varg]{txfonts}
\usepackage{graphicx}
\usepackage{natbib}
\newcommand{\isotope}[2]{${}^{#1}$#2}

\newcommand{\gcc}{\mbox {{\rm g~cm$^{-3}$}}}
\newcommand{\kms}{\mbox {{\rm km~s$^{-1}$}}}

\newcommand{\gccb}{\mbox {{\rm g~cm$^{-3}$}} } 
\newcommand{\kmsb}{\mbox {{\rm km~s$^{-1}$}} }

\begin{document} 
\nolinenumbers
\renewcommand\makeLineNumber{}

   \title{Type Ia supernovae from chemically segregated white dwarfs}

   \titlerunning{SNe Ia from chemically segregated WDs}
   
      \author{
   E. Bravo\inst{\ref{inst1}}\and J. Isern\inst{\ref{inst2},\ref{inst3},\ref{inst4}}\and L. Piersanti\inst{\ref{inst5},\ref{inst6}}
   }
   \authorrunning{E. Bravo et al.}

   \institute{
   Departamento de F\'\i sica Te\'orica y del Cosmos, Universidad de Granada, E-18071 Granada, Spain\\\email{eduardo.bravo@ugr.es}\label{inst1}
   \and
   Institut de Ci\`encies de l'Espai (ICE, CSIC), Campus UAB, C/ de Can Magrans s/n, 08193 Cerdanyola del Vall\`es, Spain\label{inst2}
   \and
   Institut d'Estudis Espacials de Catalunya (IEEC), C/ Gran Capit\`a 2-4, 08034 Barcelona, Spain\label{inst3}
   \and
    Observatori Fabra, Mathematics and Astronomy Section (RACAB),
    Rambla dels Estudis 115, planta 1a, E-08002 Barcelona, Spain\label{inst4}
   \and
   INAF-Osservatorio Astronomico d'Abruzzo, via Mentore Maggini, snc, I-64100, Teramo, Italy\label{inst5}
   \and
   INFN-Sezione di Perugia, via Pascoli, Perugia, Italy\label{inst6}
}

   \date{Received \today}

 
 \abstract
   {
   Type Ia supernovae are the outcome of the explosion of a carbon-oxygen white dwarf in a close binary system. They are thought to be the main contributors to the galactic nucleosynthesis of iron-peak elements, with important contributions to the yields of intermediate mass elements.
  Recent analyses of the phase diagram of carbon and oxygen containing impurities such as 
\isotope{22}{Ne} and \isotope{56}{Fe} in conditions relevant to white dwarf interiors suggest that 
both isotopes can partially separate when the temperature of the star is low enough to start solidifying. The purpose of this paper is to examine the impact of such a segregation on the yields of the different chemical species synthesized during explosions. 
   A one-dimensional supernova code has been used to evaluate the impact of the sedimentation assuming different degrees of chemical separation.
   It is found that the main properties of the ejecta, kinetic energy and ejected mass of \isotope{56}{Ni} do only vary slightly when the separation is taken into account. However, the yields of important isotopes that are  used as diagnostic tools such as manganese can be strongly modified.
  Furthermore, the chemical segregation studied here is able to change several indicators related to progenitor metallicity (such as the mass ratio of calcium to sulphur in the ejecta or the UV flux of the supernova) and to its mass, whether it is a Chandrasekhar-mass white dwarf or a substantially lighter one (such as the imprint of stable nickel on late-time infrared spectra or those related to the presence of radioactive nickel at the centre of the ejecta). 
  }


   \maketitle
%

\section{Introduction}
Type Ia supernovae (SNe Ia) are the outcome of the explosion of a carbon-oxygen (CO) white dwarf (WD) in a binary system. They are thought to be the main contributors to the nucleosynthesis of the galactic iron peak elements with important inputs to the synthesis of intermediate-mass isotopes, although other sources, such as core collapse supernovae, also contribute.

Observations of SN 2014J obtained with \emph{INTEGRAL} in the $\gamma$-ray domain have proved that the light curves of SNe Ia are powered by the radioactive decay of $^{56}$Ni \citep{2014Natur.512..406C,2015ApJ...812...62C,2014Sci...345.1162D,2016A&A...588A..67I}, while the near infrared observations of the innermost regions of SN 2011fe during the nebular phase have revealed the presence of important amounts of stable iron-group species such as $^ {54}$Fe and $^{58}$Ni \citep{2015MNRAS.450.2631M}, implying an important degree of neutronization. This excess of neutrons comes partially from the presence of neutron rich species inherited from the progenitor of the WD and partially from the electron captures before and during the explosive phase in those regions where density is higher than $\sim 10^9$~g~cm$^{-3}$.  If the resulting neutron excess is high enough, nuclear statistical equilibrium (NSE) matter is no longer dominated by radioactive $^{56}$Ni but by stable $^{54}$Fe and $^{58}$Ni or even more neutronized species.

The outcome of such explosions obviously depends on the abundance and distribution of the different chemical species, mass and physical state of the WD as well as on the ignition mode and the characteristics of the burning front. The scenarios that have been proposed to trigger the explosion are, loosely speaking: 
\begin{itemize}
\item
Single degenerate (SD). This scenario assumes that a CO WD accretes matter from a non-degenerate companion that can be a main sequence, a red giant star... \citep{whel73,nomo82a,hach99,han04}, or a helium star, and explodes when it reaches the Chandrasekhar's mass or when the accreted helium layer reaches a critical thickness \citep{nomo82b,woos94a,livn95}.
\item
Double degenerate (DD). This scenario assumes  two stars close enough in order to experience two episodes of common envelope evolution that lead to the formation of two WDs. If the final separation is $\lesssim 1$~R$_\odot$, the system loses angular momentum via emission of gravitational waves  at a rate that allows them to merge in less than a Hubble time \citep{iben84,webb84}.
\item
 White dwarf-white dwarf  collision (WDWD). In this scenario it is assumed that two WDs  collide and immediately ignite \citep{2009ApJ...705L.128R,lore10}. This collision can be caused by the fortuitous encounter in a dense environment, for instance the core of a globular cluster or the central regions of a galaxy, or by the interaction of a binary WD in a hierarchical triple or quadruple system via the Kozai-Lidov mechanism \citep{kush13,fang18,hame19}.
 \end{itemize}
 All these scenarios have in common the fact that the time  elapsed from the birth to the start of accretion and to the explosion of the WD can be very different depending on the parameters of the binary system. During this time the WD experiences a chemical evolution, caused by the gravitational diffusion of neutronized species and by stratification due to crystallization, that can modify the ignition density, the velocity of the burning front, and the amount of electron captures.
 
 White dwarfs with an initial mass $\lesssim 1.05$~M$_\odot$ are made of a mixture of $^{12}$C and $^{16}$O with an abundance distribution that depends on the mass and metallicity of the progenitor \citep{1997ApJ...486..413S}. Besides these two isotopes the WD contains $^{22}$Ne and other isotopes inherited from the progenitor according to its initial metallicity. In particular, $^{22}$Ne is the result of the $\alpha$-burning  of the $^{14}$N left by the hydrogen burning stage and its abundance is of the order of the sum of the initial carbon, nitrogen and oxygen abundances, X($^{22}$Ne)$\sim 0.02$ for solar metallicities.
 
 During the fluid phase, matter is composed by a mixture of carbon, oxygen and neon (CONe) \citep{1993PhRvE..48.1344O} and the chemical evolution of the star is dominated by the gravitational diffusion of the existing heavier isotopes, being $^{22}$Ne the most important reservoir of neutrons \citep{1992A&A...257..534B,2001ApJ...549L.219B,2002ApJ...580.1077D}. Self-consistent calculations have shown that $^{22}$Ne is effectively depleted in the outermost layers and that this mechanism is only efficient in massive WDs, $M\gtrsim 1$~M$\odot$ \citep{2008ApJ...677..473G,2010ApJ...719..612A,2016ApJ...823..158C} in agreement with the earlier guess by \citet{1992A&A...257..534B}. 
 
 During the process of solidification, Coulomb plasmas experience a change of miscibility and, as a consequence, the chemical profile of the stars is modified. At present it is commonly accepted that the phase diagram of solidifying CO binary mixtures has an azeotropic composition at $x_\mathrm{O}\approx 0.2$, being $x_\mathrm{O}$ the abundance of oxygen by number \citep{1988ApJ...334L..17I,1993PhRvE..48.1344O,2010PhRvL.104w1101H,2010PhRvE..81c6107M,2020A&A...640L..11B}. Since the abundance of oxygen in the major part of the star is higher than the azeotropic value, the solid mixture in equilibrium with the liquid is enriched in oxygen and, being denser, sinks towards the inner regions of the WD. The final outcome is an oxygen enrichment of the inner layers  and a depletion in the outer layers \citep{moch83,hern94,segr94,iser97,iser98,iser00,2010ApJ...717..183R,2010ApJ...716.1241S,2020A&A...640L..11B}.
 
 The distribution of trace chemical species with a neutron excess such as $^{22}$Ne and $^{56}$Fe can also be modified by the process of crystallization \citep{iser91,xu92}. If, as far as quantum effects are not relevant, the calculation of the phase diagram for Coulomb binary mixtures seems to be settled,  the case of impurities is still posing important problems \citep{2022PhR...988....1S}. 
 
 In the first attempt to guess the behaviour of $^{22}$Ne \citep{iser91} it was assumed that it was possible to approach the CONe ternary mixture by a binary one  consisting of neon and an average isotope with atomic number $\langle Z\rangle= 6x_\mathrm{C}+8x_\mathrm{O}$, being $x_\mathrm{C}$ the fraction number of carbon.  Under this hypothesis they found an azeotrope for a neon abundance in the range of 0.05-0.09 by mass. Since the abundance of neon is expected to be lower than this value, the resulting solid is poorer in neon and, being lighter than the liquid, floats and mixes with the ambient, the so called distillation process \citep{moch83}, leading to the formation of a core with the azeotropic composition and containing all the neon of the star and producing an important release of gravitational energy \citep{iser91}. Later on, assuming the same hypothesis of averaging the CO mixture, \citet{segr94} obtained an azeotropic behaviour at $x_\mathrm{Ne}=0.13$ by number.
 
 The hypothesis of an average nucleus is doubtful, at least for species with nearby atomic numbers as it was shown by \citet{1996A&A...310..485S}, who found that, if the abundance of oxygen was high, the presence of neon was irrelevant and the phase diagram was similar to a pure CO one. He also found the existence of an azeotropic point with $x_\mathrm{Ne}=0.22$, $x_\mathrm{C}=0.78$, and $x_\mathrm{O}=0.0$ that allowed the formation of a a carbon-neon shell without oxygen. Recently, \citet[][see their figure 3]{2021ApJ...911L...5B} have also found that, depending on the relative abundance of oxygen, it is possible to obtain stars with a neon rich core or a carbon-neon shell without oxygen. More precisely, depending on the initial WD composition they have found three possible configurations: a) if the liquid is rich enough in $^{22}$Ne, the solid in equilibrium is poor in neon, crystals  float and the final outcome is a solid core with a composition $(x_\mathrm{C},\,x_\mathrm{O},\,x_\mathrm{Ne})=(0.8,\,0.0,\,0.2)$ by number, surrounded by a CO mantle free of neon that crystallizes as usual; b) if the central mixture is oxygen poor, the distillation process can proceed and the final outcome is a neon rich core surrounded by a CO mantle like in case a); and c) if the composition is more or less standard, $x_\mathrm{O}=0.53,\, x_\mathrm{Ne}=0.009$ for instance, the distillation process cannot start and the solid has the composition predicted by the CO mixture without changes in the neon distribution but, since the outer layers are gradually depleted in oxygen, there is a moment at which the distillation starts  and the final outcome is a CONe core, surrounded by a carbon-neon (CNe) shell, surrounded by a CO mantle free of neon.
 
 The sedimentation of $^{56}$Fe can also play an important role \citep{xu92}. \citet{segr94}, within the hypothesis of an effective binary mixture, found a eutectic behaviour and, since the abundance of this impurity is lower than the eutectic value, the outcome is a distillation process that creates an iron rich core at the centre. This case has been recently examined by \citet{2021ApJ...919L..12C} who found that  Fe-rich crystals separate before solidification of the rest of the mixture and create an iron rich core that can be made of nearly pure iron or a carbon-oxygen-iron (COFe) alloy, depending on the exact composition of the star. Since this phenomenon occurs very early, concentric shells containing different impurities can form such as an Fe-rich core surrounded by a Ne-rich shell and both surrounded by a CO mantle, for instance.
 
 Recently, \citet{2023ApJS..268....8K} have computed the yields for a wide set of uniformly distributed $^{22}$Ne abundances in a 1.4~M$_\odot$ delayed detonation, and 1.0 and 0.8 double detonation cases. They have found that the influence is negligible for mass fractions of neon below $\lesssim 10^{-4}$ but not for abundances $\gtrsim 10^{-3}$. In any case the resulting yields have to be examined isotope by isotope.
 
Therefore, the question that arises is: what happens if neon (and iron) is not uniformly distributed because it migrates towards the central regions? This question is relevant since, depending on the system considered, the white dwarf can start to crystallize before starting the accretion process and, if this is the case, the explosion would develop within a completely different chemical structure.
 
The first attempt to analyse the impact of the $^{22}$Ne distribution on the SNe Ia nucleosynthesis\footnote{By this epoch the problem was the excess of $^{54}$Fe and $^{58}$Ni predicted by the W7 model, the most popular theoretical model of SNe Ia \citep{1984ApJ...286..644N}. } was performed by \citet{1992A&A...257..534B}, who found that if this isotope was placed at the centre its contribution to the neutron excess was negligible, thus alleviating the problem.
 
The impact of gravitational diffusion of $^{22}$Ne and sedimentation of the CO mixture, but excluding the distillation of $^{22}$Ne, on the SNe Ia yields was examined by \citet{2011A&A...526A..26B} in the Chandrasekhar and sub-Chandrasekhar scenarios, assuming spherical symmetry and neglecting convective mixing during the simmering phase. They found a negligible influence  on the supernova properties except on those depending on the ignition density and propagation of the flame. It is to be emphasized that the central abundance of \isotope{22}{Ne} assumed in the cited article is much lower than its azeotropic value.

The consequences of disrupting a WD containing a pure $^{12}$C-$^{22}$Ne layer in a WDWD collision  was analysed by \citep{2018RNAAS...2..157I,2018arXiv180908789I} who found that part of this shell could survive  providing in this way an alternative explanation to the long-lasting problem of the origin of the meteoritic Ne-E anomaly.

The time necessary to start solidification depends on the mass of the white dwarf, since the crystallization temperature depends on density, $T_\mathrm{s} \propto \rho^{1/3}$, and on the detailed chemical composition. Furthermore, it also depends on the nature of the outer envelope of the white dwarf. Those with a H-layer (DAs) cool more slowly than those lacking it (non-DAs). Typical cooling times necessary to start crystallizing are in the range of 0.2 - 6~Gyr when the mass of the white dwarf goes from 1.2 to 0.4~M$_\odot$. The typical delay time for the explosion in the SD case is in the range of 0 to 1~Gyr, while in the DD case this delay is in the 1 to 10~Gyr range, with a decaying rate $\propto t^{-1}$ \citep{2008IAUS..252..349H,2012MNRAS.426.3282M}. The typical accretion rates in the first case are in the range of $10^{-8}-10^{-7}$~M$_\odot$yr$^{-1}$ for which reason the cooling times can be enough to allow, at least, partial crystallization. In the DD case the cooling time can be arbitrarily long as well as in the case of collisions. Obviously, the casuistry is very large since it depends on the binary systems considered. Since the goal of this paper is to explore the influence of chemical sedimentation, only the extreme cases are considered.

In the next Section, there are reported the methods used in this study. In Sect.~\ref{s:22ne}, the consequences of the sedimentation of \isotope{22}{Ne} for the outcome of a SNe Ia explosion are analysed, while in Sect.~\ref{s:56fe} there are considered the consequences of bringing \isotope{56}{Fe} to the central regions of the star.
Section~\ref{s:tot} is devoted to the explosion of a WD where both isotopes, \isotope{56}{Fe} and \isotope{22}{Ne}, have migrated to the centre. Finally, in Sect.~\ref{s:conclusions}, the conclusions of the present work are exposed.

\section{Methods}

In this work, the thermonuclear supernova explosion of a WD is simulated using the one-dimensional hydrodynamic code described in detail in \citet{2019bra}. The WD is initially in hydrostatic equilibrium, isothermal, and its chemical composition is described in the following sections. The code integrates the hydrodynamic evolution using a large nuclear network including 722 isotopes \citep[the same as in][ where details can be found]{2012bra} and several thousands of nuclear reactions\footnote{All reaction rates are taken from the JINA REACLIB compilation in the version of November 6, 2008: http://groups.nscl.msu.edu/jina/reaclib/db/ \citep{2010cyb}.}, solves the Saha equations for NSE, when applicable, and calculates the neutronization rate and associated neutrino energy loses at each time step, computing the weak interaction rates on the NSE composition. The rate of the fusion reaction \isotope{12}{C}+\isotope{16}{O} was the one recommended in \citet{1988cau}. 

Concerning the explosion mechanisms, two possibilities have been considered, a detonation of a bare CO sub Chandrasekhar-mass structure, and a delayed detonation of a Chandrasekhar-mass CO WD. 
In the massive WD models, the re-homogenization by convection during simmering has not been taken into account, as justified in Sect.~\ref{chan}.

\section{\isotope{22}{Ne} sedimentation: Scenarios and models}\label{s:22ne}

 \begin{figure*}
\resizebox{\hsize}{!}{\includegraphics{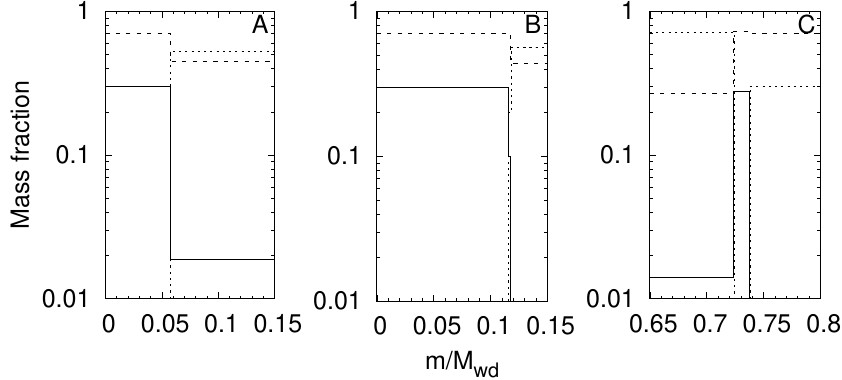}}
\caption{Details of the different enhancements on the abundance of \isotope{22}{Ne} (solid lines) and the corresponding abundances of \isotope{12}{C} (long-dashed lines) and \isotope{16}{O} (short-dashed lines) in the models reported in Table~\ref{tab1}. Case A or 'half' (left panel): Half of all \isotope{22}{Ne} concentrated in a central ball with the azeotropic composition and the other half uniformly distributed through the rest of the WD. Case B or 'full' (central panel): All \isotope{22}{Ne} concentrated in a central ball with the azeotropic composition. Case C or 'shell' (right panel): Oxygen-rich inner region with $X(^{22}\mathrm{Ne})=Z$, followed by a narrow shell rich in \isotope{22}{Ne} and depleted of oxygen, then a CO envelope with no \isotope{22}{Ne}.
}
\label{f:01}
\end{figure*}

Following \citet{2021ApJ...911L...5B}, two different scenarios for the chemical distribution of \isotope{22}{Ne} after freezing have been considered, either a central enhancement of the abundance of \isotope{22}{Ne}, if the WD metallicity is high enough (Z=0.035), cases A and B in Fig.~\ref{f:01}, or an off-centre shell heavily enriched in the same isotope containing carbon and depleted of oxygen, if the initial WD metallicity is not so high (Z=0.014), case C in Fig.~\ref{f:01}. Evidently, the number of different chemical structures can be very large, but these cases provide the frame necessary to analyse the impact of the chemical structure of the exploding WD at thermonuclear runaway.

In case B, later referred to as 'full', all the \isotope{22}{Ne} initially present in the WD is located in a central sphere with the azeotropic mass fraction, $X(^{22}\mathrm{Ne})=0.30$ and $X(^{12}\mathrm{C})=0.70$, that is the oxygen nuclei have been expelled from the centre of the WD. For a WD of mass 1.06~M$_\odot$, the full process of separation of \isotope{22}{Ne} may last of the order of 7-9~Gyrs \citep{2021ApJ...911L...5B}. Therefore, it is possible that the instability leading to a SNe Ia explosion occurs in the middle of the sedimentation process. This option is represented by case A, later referred as 'half', in which only half of \isotope{22}{Ne} nuclei have had time to settle in the central region, while the other half remains disseminated through the rest of the WD with an abundance $X(^{22}\mathrm{Ne})=0.0187$.

Case C, later referred to as 'shell', holds when the initial oxygen abundance is rather high, here it is assumed $X(^{16}\mathrm{O})=0.6$. First, an oxygen-rich and carbon-poor solid forms and migrates towards the innermost regions of the WD\footnote{According to the phase diagrams of \cite{segr94} and \cite{2020A&A...640L..11B}, the solid starts to form when the oxygen abundance grows from 0.60 to $X(^{16}\mathrm{O})=0.72$.}. In this process, the regions of the WD external to the solidified ball become enriched in carbon. When the mass fraction of carbon in these regions reach the azeotropic abundance, $X(^{12}\mathrm{C})=0.70$, a shell starts forming in which there are only carbon and \isotope{22}{Ne}, with $X(^{22}\mathrm{Ne})=0.30$. The formation of the \isotope{22}{Ne}-rich shell takes a relatively short time, of the order of a couple of Gyrs \citep{2021ApJ...911L...5B}. Finally, when all the remaining \isotope{22}{Ne} is packed in the shell, the WD is wrapped by an envelope made of carbon and oxygen. The mass coordinates of the different elements of such a chemical structure, normalized by the total WD mass, are determined by the phase diagram and the initial abundance of each isotope.

\begin{table*}
\caption{Summary of white dwarf models with different distributions of \isotope{22}{Ne}.}\label{tab1}
\centering
\begin{tabular}{rrrrrrrrr} 
\hline\hline  
\\
$M_\mathrm{WD}$ & $\rho_\mathrm{c}$ & $Z$ & \isotope{22}{Ne}$_\mathrm{core}$\tablefootmark{a} & $X(^{16}\mathrm{O})$ & $K$ & $M(^{56}\mathrm{Ni})$ & $M(^{55}\mathrm{Mn})$ & $M(^{58}\mathrm{Ni})$ \\
(M$_{\sun}$) & (g$\cdot$cm$^{-3}$) & & & & ($10^{51}$~ergs) & (M$_{\sun}$) & ($10^{-2}\mathrm{M}_{\sun}$) & ($10^{-2}\mathrm{M}_{\sun}$) \\
\hline
\\
\multicolumn{9}{l}{\underbar{Bare C-O WD detonations}} \\
1.06 & $5.4\times10^7$ & 0.035  & uniform\tablefootmark{b} & 0.50 & 1.331 & 0.652 & 0.597 & 3.94 \\
1.06 & $5.7\times10^7$ & 0.035  & half\tablefootmark{c}    & 0.50 & 1.328 & 0.641 & 0.548 & 4.70 \\
1.06 & $5.9\times10^7$ & 0.035  & full\tablefootmark{d}    & 0.50 & 1.330 & 0.635 & 0.238 & 6.36 \\
1.06 & $5.3\times10^7$ & 0.014  & uniform\tablefootmark{b} & 0.60 & 1.266 & 0.678 & 0.390 & 1.29 \\
1.06 & $5.3\times10^7$ & 0.014  & shell\tablefootmark{e}    & 0.60 & 1.285 & 0.682 & 0.382 & 1.30 \\
\\
\multicolumn{9}{l}{\underbar{Chandrasekhar-mass WD delayed detonations}} \\
1.38 & $5.0\times10^9$ & 0.035  & uniform\tablefootmark{b} & 0.50 & 1.382 & 0.601 & 1.58 & 3.78 \\
1.37 & $5.0\times10^9$ & 0.035  & half\tablefootmark{c} & 0.50 & 1.369 & 0.629 & 1.22 & 2.56 \\
1.37 & $5.0\times10^9$ & 0.035  & full\tablefootmark{d} & 0.50 & 1.355 & 0.625 & 0.407 & 2.78 \\
1.38 & $6.9\times10^9$ & 0.035  & full\tablefootmark{f} & 0.50 & 1.338 & 0.629 & 0.414 & 1.95 \\
1.38 & $5.0\times10^9$ & 0.014  & uniform\tablefootmark{b} & 0.60 & 1.283 & 0.603 & 1.12 & 2.71 \\
1.38 & $5.0\times10^9$ & 0.014  & shell\tablefootmark{e} & 0.60 & 1.287 & 0.540 & 1.05 & 2.73 \\
\\
\hline
\end{tabular}
\tablefoot{
\tablefoottext{a}{Degree of concentration of \isotope{22}{Ne} in a central core with azeotropic abundance, $X(^{22}\mathrm{Ne})=0.30$.}
\tablefoottext{b}{Uniform composition through the whole WD, with $X(^{22}\mathrm{Ne})=Z$ and $X(^{12}\mathrm{C})=1-X(^{22}\mathrm{Ne})-X(^{16}\mathrm{O})$.}
\tablefoottext{c}{Half of all \isotope{22}{Ne} concentrated in a central ball with the azeotropic composition; the other half uniformly distributed through the rest of the WD (case A in Fig.~\ref{f:01}).}
\tablefoottext{d}{All \isotope{22}{Ne} concentrated in a central ball with the azeotropic composition (case B in Fig.~\ref{f:01}).}
\tablefoottext{e}{Enhanced \isotope{22}{Ne} abundance in a narrow off-centre shell (case C in Fig.~\ref{f:01}).}
\tablefoottext{f}{Same as case B in Fig.~\ref{f:01} but with the same mass as the WD with uniform composition.}
}
\end{table*}

The list of explosion models is given in Table~\ref{tab1}, where $M_\mathrm{WD}$ is the mass of the WD, $\rho_\mathrm{c}$ its central density, $Z$ is the progenitor metallicity, which equals the overall abundance of \isotope{22}{Ne} in the WD, and $X(^{16}\mathrm{O})$ is the abundance of \isotope{16}{O}. The radial profile of the distribution of \isotope{22}{Ne} is given as \isotope{22}{Ne}$_\mathrm{core}$: it may be either uniform, half, full, or shell. The uniform case, in which the chemical composition of the WD is homogeneous, is the reference for comparisons with heterogeneous models. The main result of the explosion is that the kinetic energy, $K$, and the ejected mass of \isotope{56}{Ni}, $M(^{56}\mathrm{Ni})$  are almost independent  of the distribution of \isotope{22}{Ne}, while the ejected masses of \isotope{55}{Mn} and \isotope{58}{Ni} strongly depend on the chemical structure. These data are given at a time of 100~s after thermal runaway.

\subsection{Bare CO WD detonation}

In this section, there are shown the results of the explosion of a sub Chandrasekhar-mass WD with the different distributions of \isotope{22}{Ne} aforesaid. The WD mass is $1.06$~M$_\sun$ and the helium cap responsible for the instability leading to a central detonation is left out of the computation \citep{2010sim}. Once a detonation is initiated in the helium layer, a detonation may be started in the CO core either directly at the interface with the helium envelope or by convergence of an inward moving shock wave near the centre of the WD. In one-dimensional simulations with central carbon mass fraction similar or even lower than assumed here, a detonation is obtained \citep[e.g.][]{2011woo}. In this work, the hydrodynamic simulation begins with the central detonation of carbon.

\subsubsection{Central enhancement of the \isotope{22}{Ne} abundance}
\begin{figure}
\resizebox{\hsize}{!}{\includegraphics{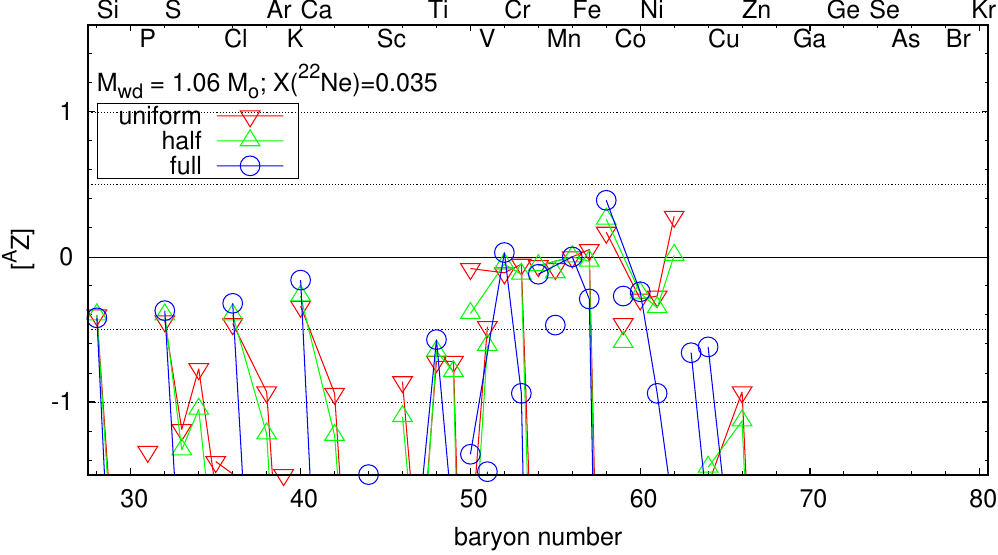}}
\caption{Nucleosynthesis of sub Chandrasekhar-mass bare CO models with $M_\mathrm{WD}=1.06$~M$_\sun$ and $Z=0.035$ with different distributions of \isotope{22}{Ne}. The abundance of each isotope is given with reference to \isotope{56}{Fe} and normalized to Solar System abundances: 
$
\left[^{A}Z\right] = \mathrm{log}_{10}\left[X\left(^{A}Z\right)/X\left(^{56}\mathrm{Fe}\right)\right]-\mathrm{log}_{10}\left[X\left(^{A}Z\right)/X\left(^{56}\mathrm{Fe}\right)\right]_\odot\,,
$
\noindent where $X$ stands for mass fraction. All the unstable species have been allowed to decay to their stable isobars. 
}
\label{f:a}
\end{figure}

Figure~\ref{f:a} shows the nucleosynthesis output of the explosion of sub Chandrasekhar-mass bare CO models with $M_\mathrm{WD}=1.06$~M$_\sun$ and $Z=0.035$ with different distributions of \isotope{22}{Ne}. Only a few under-produced isotopes (with respect to \isotope{56}{Fe} and the Solar System abundances) are affected by the distribution of \isotope{22}{Ne}, where uniform and half yields are very similar, while full yields are noticeably different.

Comparing the yields of uniform and half models, the most relevant differences are: \isotope{64}{Zn} and \isotope{66}{Zn} differ by $\sim30-40$\% between uniform and half models, but the first is favoured after partial \isotope{22}{Ne} distillation and the second in the homogeneous case, \isotope{58}{Ni} is more abundant by $\sim19$\% in the half model, while \isotope{50}{Cr} and \isotope{62}{Ni} decrease by $\sim50$\% in the same model.

With respect to uniform versus full models, the most relevant differences after complete \isotope{22}{Ne} distillation are: increments of \isotope{63}{Cu} by a factor 51, of \isotope{64}{Zn} by a factor 9, of \isotope{59}{Co} by $\sim50$\%, and of \isotope{58}{Ni} by $\sim60$\%, and decrements of \isotope{62}{Ni} and \isotope{66}{Zn} by two orders of magnitude, \isotope{50}{Cr} and \isotope{53}{Cr} by factors 20 and 8, respectively, \isotope{61}{Ni} by a factor $\sim5$, \isotope{57}{Fe} by a factor 2, and many other by smaller amounts. 

The yields of \isotope{55}{Mn} are also affected, as they decrease after complete \isotope{22}{Ne} distillation by a factor 2.5 with respect to the uniform model. \isotope{55}{Mn} is the only stable isotope of manganese, which is an important product of SNe Ia and relevant for chemical evolution \citep[e.g.][]{2020eit}, although this decrease is not going to change the conclusions of the cited studies. 

To summarise, the total or partial concentration of \isotope{22}{Ne} in the central regions of a WD favour the synthesis of cobalt and copper in a SNe Ia explosion, while it disfavours manganese and chromium. The sedimentation of \isotope{22}{Ne} also changes the isotopic composition of nickel. In this respect, it is interesting to note that the amount of stable nickel isotopes synthesized in a SNe Ia has been suggested to correlate with the mass of the exploding WD, providing a discriminant between a Chandrasekhar-mass progenitor versus a sub Chandrasekhar-mass one \citep{2022blo}. However, \isotope{22}{Ne} sedimentation can change the trend. For instance, in the full model the yield of \isotope{58}{Ni} is 0.064~M$_\odot$, which is higher than in any of the Chandrasekhar-mass models presented in Sect.~\ref{chan} \citep[see also Fig.~5 and Table~A.1 in][]{2022blo}.
 
It is also remarkable that there is an increase in the mass ratio of calcium with respect to sulphur of the order of 28\% in the full model as compared with the uniform model, as a consequence of the disappearance of the neutron excess carried by \isotope{22}{Ne} in the layers where these species are synthesized. This effect, even modest in magnitude\footnote{\cite{2017mar} measure calcium to sulphur mass ratios varying as much as 69\% between different SNe Ia remnants.}, may be relevant enough and help with matching X-ray spectral data concerning the abundances of these elements in SNe Ia remnants \citep{2014de,2017mar}, as it can make the explosion of a WD with a super-solar metallicity progenitor appear as due to a low-metallicity one.

\begin{figure}
\resizebox{\hsize}{!}{\includegraphics{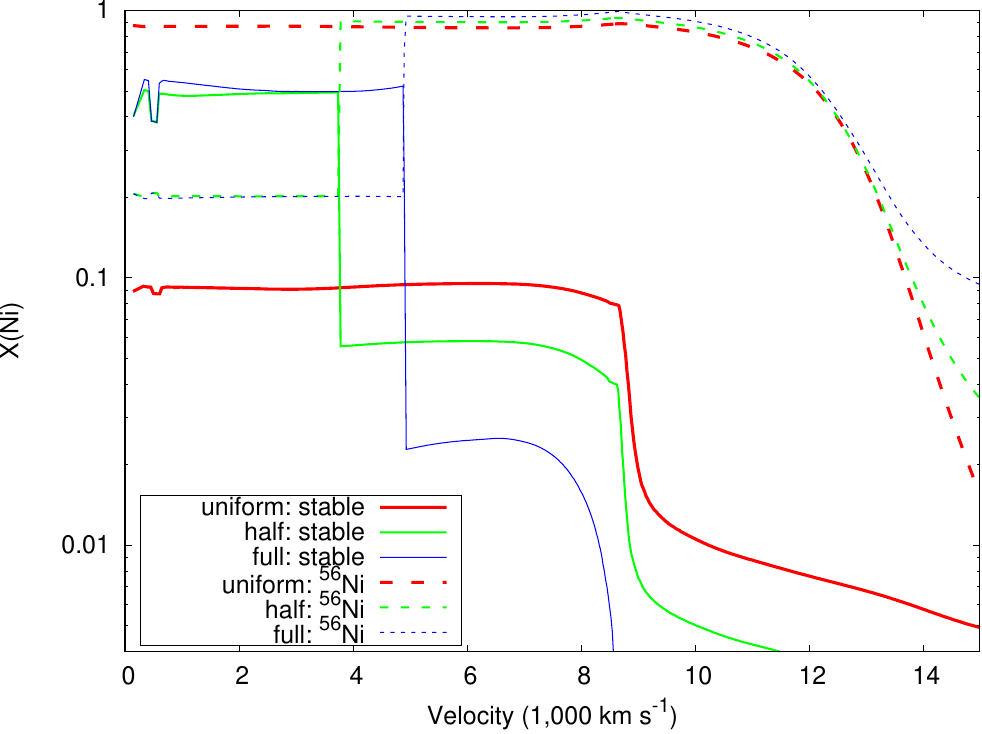}}
\caption{Profile of \isotope{56}{Ni} and stable nickel in the inner regions of the ejecta of the models with $M_\mathrm{WD}=1.06$~M$_\sun$ and $Z=0.035$ with different distributions of \isotope{22}{Ne} (see Table~\ref{tab1} for details). The maximum velocity shown in the plot belongs to a Lagrangian mass coordinate $\sim0.91$~M$_\sun$.
}
\label{f:10}
\end{figure}

Besides the nucleosynthetic outcome of SNe Ia, the redistribution of the neutron excess of the WD prior to its explosion may have some more directly observable consequences. One of them is related to the profile of stable and radioactive nickel isotopes (Fig.~\ref{f:10}). In contrast with the homogeneous case, in both the half and the full models there is a high abundance of stable nickel in the layers moving slower than 4,000-5,000~\kms, while \isotope{56}{Ni} is absent in the same regions. In SN 2003du, for instance, the shape of infrared lines due to $[$Fe II$]$ suggests that the central volume of the ejecta up to an expanding velocity of 3,000~\kmsb is mostly made of non-radioactive iron-group isotopes, and is surrounded by a region of radioactive \isotope{56}{Ni} up to 10,000~\kmsb \citep{2004hoe}. Such chemical structure has been attributed to burning at high density as expected in Chandrasekhar-mass progenitors. However, as shown in Fig.~\ref{f:10}, the same structure can be mimicked by a sub Chandrasekhar-mass progenitor in which \isotope{22}{Ne} has been total or partially concentrated in the centre of the WD prior to the supernova explosion. 

\subsubsection{
Explosion after \isotope{22}{Ne} distillation in a solar metallicity WD
}

\begin{figure}
\resizebox{\hsize}{!}{\includegraphics{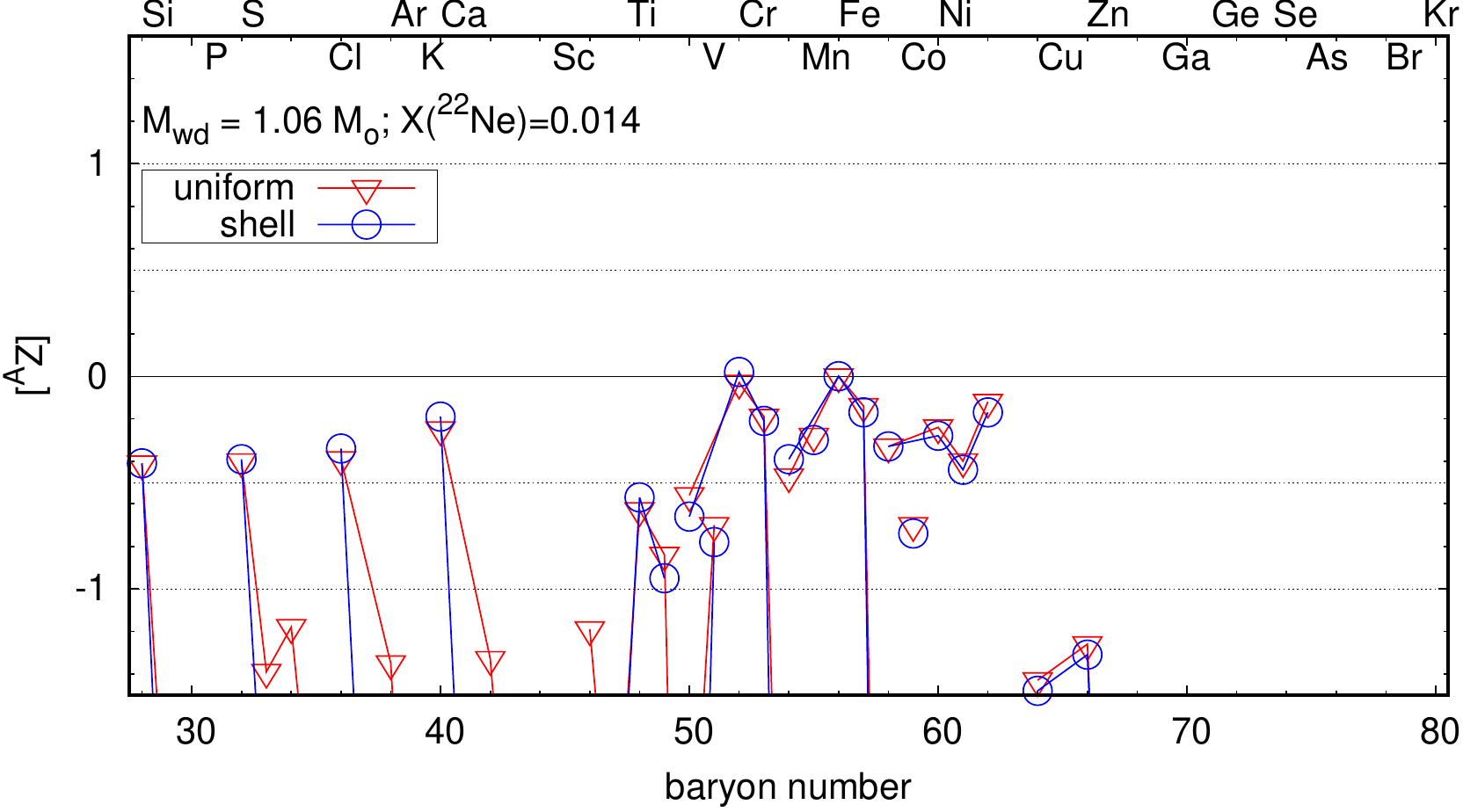}}
\caption{Same as Fig.~\ref{f:a}, but for the sub Chandrasekhar-mass models with $M_\mathrm{WD}=1.06$~M$_\sun$ and $Z=0.014$.
}
\label{f:b}
\end{figure}

Figure~\ref{f:b} shows the final yields of the explosions with $M_\mathrm{WD}=1.06$~M$_\sun$, $Z=0.014$ and different distributions of \isotope{22}{Ne}, either homogeneous or an oxygen-rich core followed by a narrow off-centre shell with the azeotropic composition of carbon and neon surrounded by a CO envelope (the shell model). 

As expected, and evidenced by Fig.~\ref{f:b}, decreasing the progenitor metallicity reduces the effect of redistributing the \isotope{22}{Ne} through the WD. In the shell model, the most noticeable change affects \isotope{54}{Fe}, whose yield increases with respect to the uniform model by 21\%, while the final abundances of \isotope{40}{Ca} and \isotope{52}{Cr} increase by 15\%. On the other hand, the yields of many neutron-rich isotopes of intermediate-mass elements decrease in the shell model, but their abundance was already quite low in the uniform model. The only isotope with a non-negligible yield in the uniform model that is negatively affected by the redistribution of neutron excess in the shell model is \isotope{50}{Cr}, whose abundance decreases by 20\%.

\begin{figure}
\resizebox{\hsize}{!}{\includegraphics{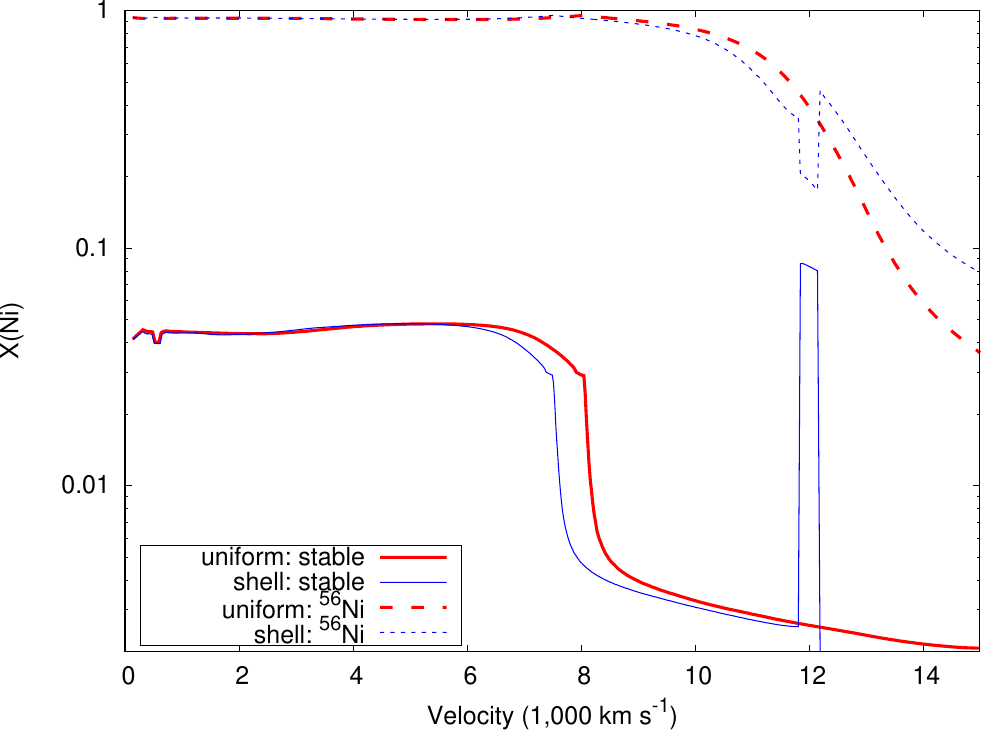}}
\caption{Profile of \isotope{56}{Ni} and stable nickel in the inner regions of the ejecta of the models with $M_\mathrm{WD}=1.06$~M$_\sun$ and $Z=0.014$ with different distributions of \isotope{22}{Ne} (see Table~\ref{tab1} for details). 
}
\label{f:11}
\end{figure}

Figure~\ref{f:11} shows the profiles of the final mass fraction of stable nickel isotopes and that of \isotope{56}{Ni} (alike in Fig.~\ref{f:10}). The concentration of \isotope{22}{Ne} and, consequently, of neutron excess in an off-centre shell leaves an almost imperceptible imprint but for the abundance spike of stable nickel, visible at an expansion velocity of $\sim12,000$~\kmsb in the shell model. Such an enhancement of nickel abundance in a narrow shell may be hard to detect in actual supernovae.

\subsection{Chandrasekhar-mass white dwarfs}\label{chan}

In this section, it is assumed that the explosion is triggered by a delayed detonation starting as a deflagration in a Chandrasekhar-mass WD with a central density of $\rho_\mathrm{c}=5\times10^9$~\gccb \citep{2022pie} in the uniform and in the chemically segregated cases.

A previous question to address is if the chemically segregated structure is able to survive the accretion and simmering phases that precede thermal runaway. The simmering phase starts when the temperature is high enough so that cooling by thermal neutrinos cannot compensate the nuclear energy release rate, and ends with a thermonuclear runaway, and it is characterized by an efficient convective mixing of the innermost regions of the WD. 
As shown by \cite{2022pie}, convection during the simmering phase may be restricted to the central $\sim0.11$~M$_\odot$. Therefore, the chemical structure of the initial models shown in Fig.~\ref{f:01} may be preserved but, perhaps, for case A. In case C, it will be preserved for sure\footnote{The density of the CNe shell is higher than that of the supporting CONe layer meaning that, if it melts as a consequence of accretion, the CNe shell will progressively mix with inner layers. 
}.

\subsubsection{Central enhancement of the \isotope{22}{Ne} abundance}

\begin{figure}
\resizebox{\hsize}{!}{\includegraphics{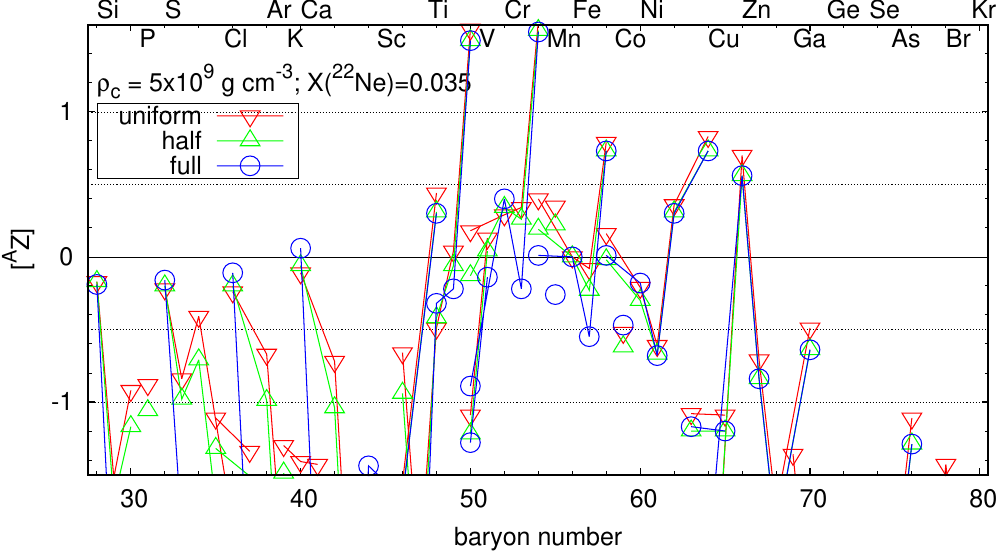}}
\caption{Same as Fig.~\ref{f:a}, but for the Chandrasekhar-mass models with $\rho_\mathrm{c}=5\times10^9$~\gccb and $Z=0.035$.
}
\label{f:c}
\end{figure}

Figure~\ref{f:c} shows the nucleosynthetic results in the uniform, full, and half cases. The impact of the sedimentation of \isotope{22}{Ne} at the centre of the WD is even milder than in the sub Chandrasekhar-mass models. This is because the centre of the massive WD experiences huge electron captures during the first second of the explosion, which erases most of the traces of the previous presence of a neutron excess there. When comparing the half and the uniform models, the most relevant change in the abundances affects \isotope{50}{Cr}, whose yield is reduced in the half model by a factor two, while those of \isotope{54}{Fe} and \isotope{58}{Ni} are reduced by $\sim35\%$.

On the other hand, in the full model the only substantial yield increase with respect to the uniform model affects \isotope{52}{Cr}, whose abundance increases by 30\%. By contrast, the production of many isotopes becomes reduced when \isotope{22}{Ne} is fully concentrated at the centre of the WD. The most affected, besides many neutron-rich isotopes of intermediate-mass elements, are: \isotope{50}{Cr} by an order of magnitude, \isotope{55}{Mn} and \isotope{53}{Cr} by a factor four, \isotope{54}{Cr} and \isotope{57}{Fe} by 60\%, \isotope{51}{V} and \isotope{49}{Ti} by 40\%, \isotope{58}{Ni}, \isotope{66}{Zn} and \isotope{67}{Zn} by a 25\%, and so on. 

As in the sub Chandrasekhar-mass WD models, the \isotope{40}{Ca} to \isotope{32}{S} mass ratio increases in the full model by 30\%, with potential implications for the interpretation of X-ray spectra of SNe Ia remnants.

\subsubsection{Influence of the conditions at thermal runaway on the final yields}

The presence of \isotope{22}{Ne} in the central regions implies that, for the same mass of the WD, the central density has to be higher to reach the hydrostatic equilibrium, and the importance of this effect increases when approaching to the Chandrasekhar mass. Therefore,
when comparing the effects of sedimentation with respect to the uniform model, one can choose to keep fixed the central density at thermal runaway, as in Fig.~\ref{f:c} and reported in the previous paragraphs, or to keep the WD mass constant, as it was done in the sub Chandrasekhar-mass scenario. As reported in Table~\ref{tab1}, the full model with a mass equal to that of the uniform model has a central density of $6.9\times10^9$~\gcc. Still, the kinetic energy and ejected mass of \isotope{56}{Ni} are quite similar to those in the uniform model with a central density of $5.0\times10^9$~\gcc. 

\begin{figure}
\resizebox{\hsize}{!}{\includegraphics{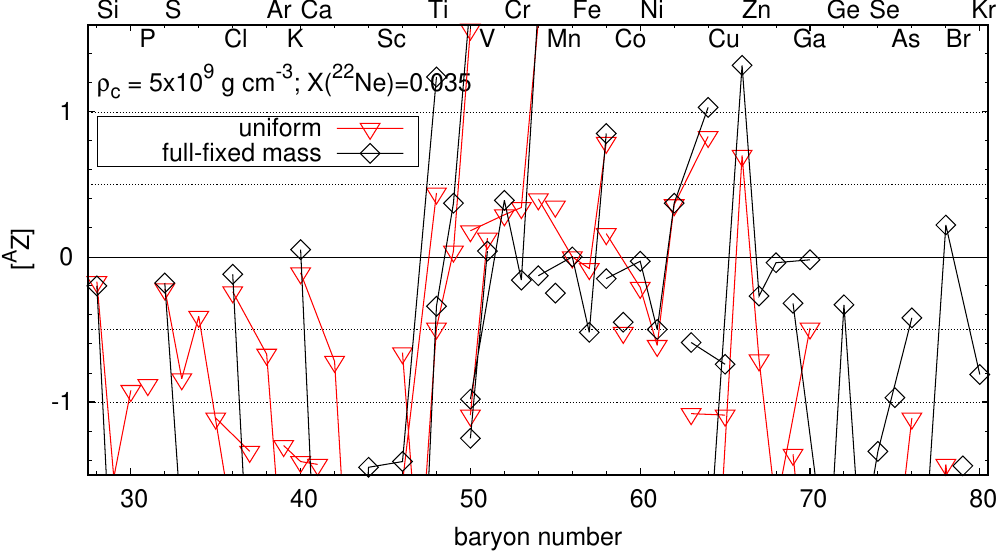}}
\caption{Same as Fig.~\ref{f:c}, but comparing the uniform composition model with case B maintaining the same mass of the WD.
}
\label{f:d}
\end{figure}

In Fig.~\ref{f:d}, the yields of the uniform and full models with the same mass are compared. Due to the huge difference in central density at thermal runaway, the isotopic yields of both models are quite different. In the full model, selenium, germanium, gallium, zinc and copper abundances increase from a factor of a few up to two orders of magnitude. On a more modest level, the synthesis of the most overproduced isotopes, \isotope{54}{Cr} and \isotope{50}{Ti}, is favoured by the higher density of the full model, and their abundances increase by 36\% and 67\%, respectively, with respect to the uniform model. On the other hand, there are strong deficits of many isotopes of intermediate-mass elements, while the yields of \isotope{58}{Ni}, \isotope{54}{Fe} and \isotope{55}{Mn} are reduced by factors two, three and four, respectively.

\subsubsection{Explosion after \isotope{22}{Ne} distillation in a solar metallicity WD}

\begin{figure}
\resizebox{\hsize}{!}{\includegraphics{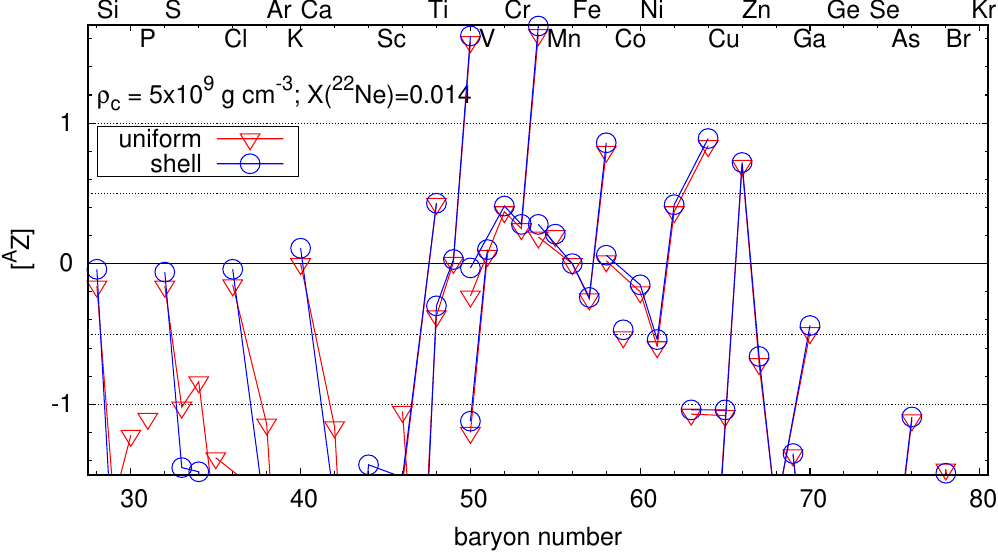}}
\caption{Same as Fig.~\ref{f:c}, but for the Chandrasekhar-mass models with $\rho_\mathrm{c}=5\times10^9$~\gccb and $Z=0.014$.
}
\label{f:e}
\end{figure}

Concerning the shell model, the impact on the nucleosynthesis of the Chandrasekhar-mass explosions is similar to that on the sub Chandrasekhar-mass ones, meaning that the quantitative impact is quite modest except for some neutron-rich isotopes of intermediate-mass nuclei (Fig.~\ref{f:e}). Among the few species whose yields are modified, the most important is \isotope{50}{Cr}, whose abundance rises by 42\%, and \isotope{46}{Ti}, whose yield decreases by a factor three. Nevertheless, as can be seen in Table~\ref{tab1}, it is noticeable that the yield of \isotope{56}{Ni} decreases by $\sim10\%$ when \isotope{22}{Ne} is concentrated in an off-centre shell on a massive WD.

\section{\isotope{56}{Fe} sedimentation: Scenarios and models}\label{s:56fe}

As already stated in the Introduction, an iron-enhanced solid phase can form thanks to the high atomic number of \isotope{56}{Fe} when the temperature is still too hot for a CO mixture to solidify. The solid can either be pure iron or a COFe alloy in which the iron number abundance is $\sim15\%$, that is an iron mass fraction $X(^{56}\mathrm{Fe})\sim0.4$. In the case of pure iron, a solid ball of $\sim0.001$ solar masses forms in the centre of the WD. If the solid has the composition of the alloy, the size of the central region enriched in iron is 2.5 times larger \citep{2016ApJ...823..158C}.

Since it is not possible to extract nuclear energy from iron, a pure iron ball does not favour the development of a thermonuclear supernova explosion and it is necessary to explore the effects of different degrees of iron enrichment in the center of a WD. For these experiments, iron mass fractions from $X(^{56}\mathrm{Fe})\sim10^{-3}$, which corresponds to a Solar System abundance of \isotope{56}{Fe}, to $X(^{56}\mathrm{Fe})\sim0.7$ have been used. The last value, and intermediate ones, are meant to represent either the COFe alloy mentioned earlier or an intermediate phase in the formation of the pure iron core at the centre of a WD. In the models with $X(^{56}\mathrm{Fe})>10^{-3}$, that is all models but the homogeneous one, it is assumed that all iron is concentrated in a central ball, albeit the size of the ball changes as a function of the central value of $X(^{56}\mathrm{Fe})$, such that the total iron abundance in the WD remains the same in all cases. 

\subsection{Thermonuclear burning waves in COFe mixtures}

Since the properties of burning waves can be affected by the high abundances of iron assumed here, it is necessary to start exploring the feasibility of detonation waves and the structure and velocity of conductive flames in COFe mixtures. In the models reported in Sect.~\ref{s:scfe}, a sub Chandrasekhar-mass CO WD detonates as a consequence of a converging wave that ignites matter near the centre of the star at a density of the order of a few times $10^7$~\gcc. On the other hand, in the models reported in Sect.~\ref{s:cfe} conductive flames propagate through the centre of a massive WD after experiencing a simmering phase, that is during the first stages of the deflagrative phase of a delayed detonation, when the WD central density is of the order of $\rho_\mathrm{c}=5\times10^9$~\gccb  \citep{2022pie}. In the delayed detonation models, the detonation starts outside the zone enriched in iron. 

\begin{figure*}
\centering
\resizebox{\hsize}{!}{\includegraphics{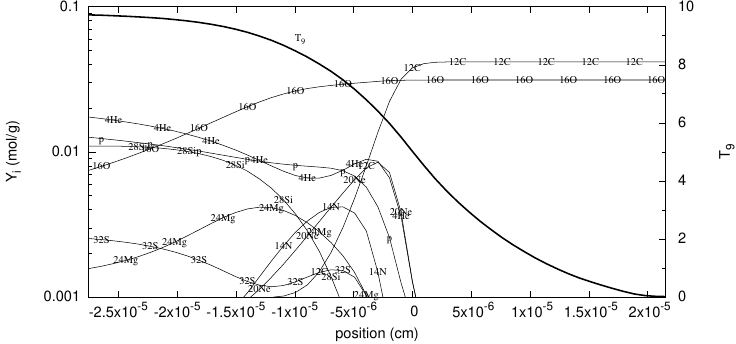}}
\hfill
\resizebox{\hsize}{!}{\includegraphics{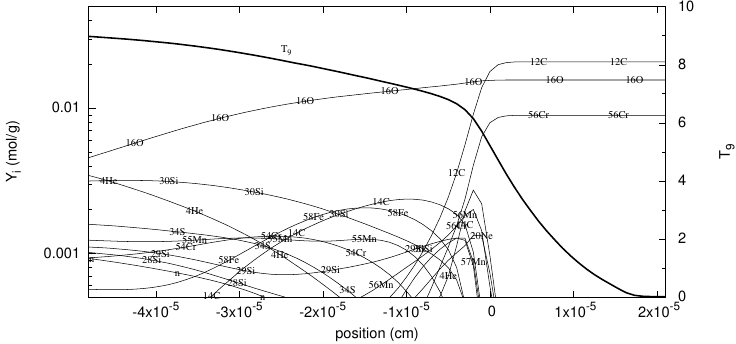}}
\caption{
Chemical structure of a conductive flame in 50\%-50\% \isotope{12}{C}-\isotope{16}{O} (top) and 50\%-25\%-25\% \isotope{56}{Cr}-\isotope{12}{C}-\isotope{16}{O} (bottom) mixtures at a density of $5\times10^9$~\gcc. The temperature is plotted with a thick line. The initial width of the shells was $10^{-6}$~cm.
}
\label{f:1}
\end{figure*}

\begin{figure*}
\centering
\resizebox{\hsize}{!}{\includegraphics{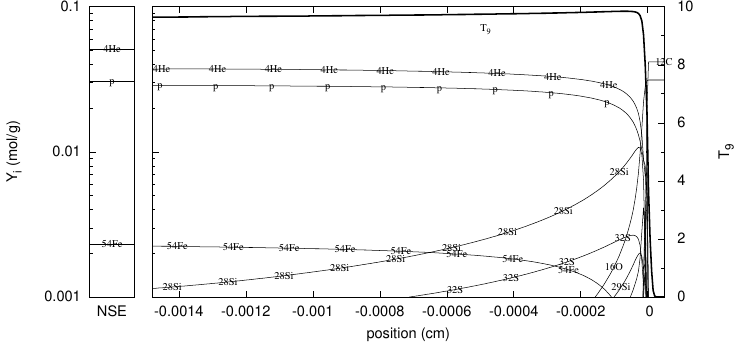}}
\hfill
\resizebox{\hsize}{!}{\includegraphics{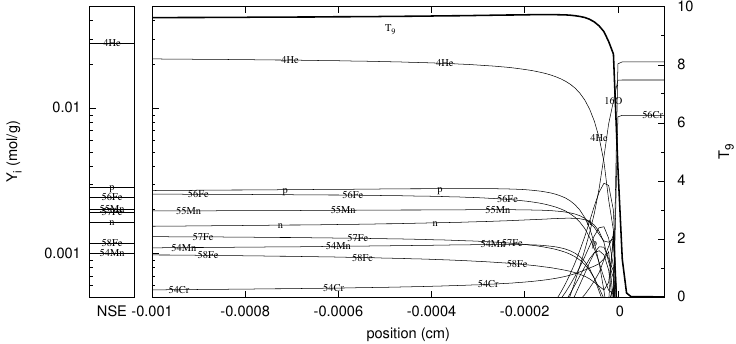}}
\caption{
Same as Fig.~\ref{f:1} but on a longer length scale. At the left, there are shown the molar abundances belonging to nuclear statistical equilibrium. For this flame simulations, the initial width of the shells was $5\times10^{-6}$~cm (top) and $10^{-5}$~cm (bottom), while the respective electron mole numbers are 0.500 and 0.464~mol~g$^{-1}$.
}
\label{f:2}
\end{figure*}


\subsubsection{Detonation}

The study of the feasibility of a detonation in iron-rich matter follows a well known procedure \citep[e.g.][]{1997nie} consisting in solving the hydrodynamic and nuclear kinetic equations at (initial) constant density, with high spatial and temporal resolution. A homogeneous sphere made of a mixture of carbon, oxygen and iron is endowed with a shallow temperature gradient at the centre (the hot spot), such that shells burn one after the other in sequence and a pressure wave grows into the unburnt matter. If the hot spot is large enough, the pressure wave transitions into a self-sustained detonation. The \isotope{56}{Fe} mass fraction has been given different values: $X(^{56}\mathrm{Fe})=0$, 0.35, 0.50, and 0.70, while both \isotope{12}{C} and \isotope{16}{O} mass fractions are equal to $0.5\left[1-X(^{56}\mathrm{Fe})\right]$.

The code used in this study is the same hydrodynamic code used in the explosion models with gravitation turned off. In the present detonation calculations, the density is $\rho=5\times10^7$~\gccb and the maximum temperature in the centre of the detonating sphere is $T_\mathrm{max}=3.2\times10^9$~K but for the composition with $X(^{56}\mathrm{Fe})=0.70$, in which case it is $T_\mathrm{max}=4\times10^9$~K. The minimum size of a hotspot able to detonate a COFe sphere has been computed as well with a lower maximum temperature, $T_\mathrm{max}=2.8\times10^9$~K, without substantial variations in the results. 

If the iron concentration stays below $X(^{56}\mathrm{Fe})\simeq0.5$, the size of the hot spot needed to start a detonation at $\rho=5\times10^7$~\gccb lies between a few meters and 1~km, even though the precise values are somewhat dependent on the details of the thermal profile imprinted initially \citep{2009sei}. Since the radius of the iron-rich sphere in a solar-metallicity WD is $200-300$~km, a detonation is a likely result even when half of the mass of the central sphere is iron. 

However, if the iron abundance in the central sphere is as high as $X(^{56}\mathrm{Fe})=0.70$, a detonation is not obtained even for hot spot sizes as long as 1,000~km, that is more than 3-4 times the size of the iron-rich region in a solar-metallicity WD, or more than 100 times its mass. Therefore, if the central iron mass fraction is high enough, a detonation can only be initiated outside the central iron-rich ball.

\subsubsection{Conductive flame}\label{ss:cf}

As in the studies of detonating spheres, the properties of conductive flames with different iron abundances have been determined by solving the hydrodynamic equations coupled to the nuclear kinetic equations with the same code used for the supernova explosions with gravity turned off \citep[such as, for instance, the micro-zone hydrocode method described in][]{1992tim}. The flame structure has been computed for $\rho=5\times10^9$~\gcc, at which density \isotope{56}{Fe} is unstable to a double electron capture. Therefore, all iron has been transformed into \isotope{56}{Cr} before starting the hydrodynamic calculations. As in the previous section, the flame structure has been calculated for different values of the \isotope{56}{Cr} mass fraction: $X(^{56}\mathrm{Cr})=0$, 0.35, 0.50, and 0.70, while both \isotope{12}{C} and \isotope{16}{O} mass fractions are equal to $0.5\left[1-X(^{56}\mathrm{Cr})\right]$.

The flame velocity decreases with increasing \isotope{56}{Cr} abundance, from 140~\kmsb when $X(^{56}\mathrm{Cr})=0$ to 20~\kmsb when $X(^{56}\mathrm{Cr})=0.70$. At the assumed density, the size of the iron-rich ball in a solar-metallicity WD is of the order of 70~km, which means that it can be consumed by the flame in less than one second if the abundance of \isotope{56}{Cr} is $\lesssim0.5$ (flame velocity $\gtrsim68$~\kms). 

The thermal and chemical structure of conductive flames at a density of $5\times10^9$~\gccb are shown in Figs.~\ref{f:1} and \ref{f:2} for two different cases, either matter composed initially just of \isotope{12}{C} and \isotope{16}{O}, or a mixture of both isotopes and \isotope{56}{Cr}, with $X(^{56}\mathrm{Cr})=0.50$.

In the first case (top panel of Fig.~\ref{f:1}), the chemical transmutation starts with the fusion reaction \isotope{12}{C}+\isotope{12}{C} releasing both protons and alphas in almost equal numbers, together with \isotope{20}{Ne} and \isotope{23}{Na}, followed by $^{23}\mathrm{Na}({\mathrm p},\alpha)^{20}\mathrm{Ne}$. Shortly after, the \isotope{20}{Ne} abundance decreases due to $^{20}\mathrm{Ne}(\alpha,{\mathrm p})^{23}\mathrm{Na}$ and $^{20}\mathrm{Ne}({\mathrm p},\alpha)^{17}\mathrm{O}$, and the $\alpha$ abundance equilibrates with those of \isotope{14}{C}, \isotope{17}{O} and neutrons, while the abundance of protons stabilizes through multiple reactions with intermediate-mass elements and carbon and oxygen isotopes. Thereafter, \isotope{28}{Si} and other intermediate-mass elements are synthesized following the oxygen-fusion reaction. Finally (top panel of Fig.~\ref{f:2}), the abundance of \isotope{54}{Fe} increases and approaches its value at NSE. 

In the second case (bottom panel of Fig.~\ref{f:1}), the high initial abundance of \isotope{56}{Cr} is a trap for protons through the reaction $^{56}\mathrm{Cr}({\mathrm p},{\mathrm n})^{56}\mathrm{Mn}$, which prevents protons from contributing significantly to the building of intermediate-mass elements. In this case, there is never a large amount of silicon or sulphur isotopes. Most species from the iron-group are the result of $({\mathrm p},{\mathrm n})$ reactions with some contribution from $(\gamma,{\mathrm n})$ disintegrations, while intermediate-mass elements are built on the basis of $\alpha$-captures. Starting with a high abundance of \isotope{56}{Cr} is quite efficient for achieving NSE, as can be seen comparing both panels in Fig.~\ref{f:2}: the molar fractions of the most abundant isotopes change only slightly beyond $\sim2\times10^{-4}$~cm behind the flame front, while in the top panel there can be seen a sizeable variation in the abundance of \isotope{28}{Si} after the front has advanced $10^{-3}$~cm. In spite of the highest efficiency, the width of the flame front (as defined by the destruction of both \isotope{12}{C} and \isotope{16}{O}) is longer in the second case than in the first case. 

\begin{table*}
\caption{Summary of SNe Ia models with different central enhancements of \isotope{56}{Fe} (sub Chandrasekhar-mass WDs) or \isotope{56}{Cr} (Chandrasekhar-mass WDs).}\label{tab3}
\centering
\begin{tabular}{rrrrrrr} 
\hline\hline  
$M_\mathrm{WD}$ & $\rho_\mathrm{c}$ & $X_\mathrm{c}(A=56)$\tablefootmark{a} & $K$ & $M(^{56}\mathrm{Ni})$ & $M(^{55}\mathrm{Mn})$ & $M(^{58}\mathrm{Ni})$ \\
(M$_{\sun}$) & (\gcc) & & ($10^{51}$~ergs) & (M$_{\sun}$) & ($10^{-2}\mathrm{M}_{\sun}$) & ($10^{-2}\mathrm{M}_{\sun}$) \\
\hline
\multicolumn{7}{l}{\underbar{Bare C-O WD detonations}} \\
1.06 & $4.8\times10^7$ & $10^{-3}$ & 1.319 & 0.655 & 0.410 & 1.35 \\
1.06 & $5.0\times10^7$ & 0.35 & 1.324 & 0.656 & 0.402 & 1.39 \\
1.06 & $5.1\times10^7$ & 0.50 & 1.324 & 0.656 & 0.399 & 1.34 \\
1.06 & $5.2\times10^7$ & 0.70\tablefootmark{b} & 1.323 & 0.656 & 0.395 & 1.27 \\
\multicolumn{7}{l}{\underbar{Chandrasekhar-mass WD delayed detonations}} \\
1.37 & $3.5\times10^9$ & $10^{-3}$ & 1.410 & 0.669 & 1.05 & 3.08 \\
1.37 & $3.5\times10^9$ & 0.35 & 1.418 & 0.671 & 1.07 & 2.87 \\
1.37 & $3.5\times10^9$ & 0.70 & 1.421 & 0.675 & 1.01 & 2.97 \\
1.38 & $5.0\times10^9$ & $10^{-3}$ & 1.389 & 0.640 & 1.16 & 2.90 \\
1.38 & $5.0\times10^9$ & 0.35 & 1.401 & 0.651 & 1.12 & 2.85 \\
1.38 & $5.0\times10^9$ & 0.70 & 1.405 & 0.655 & 1.14 & 2.79 \\
\hline
\end{tabular}
\tablefoot{All models assumed a progenitor metallicity $Z=0.014$. 
\tablefoottext{a}{Mass fraction of the stable isotope with $A=56$ in the WD centre: \isotope{56}{Fe} in sub Chandrasekhar-mass WDs; \isotope{56}{Cr} in Chandrasekhar-mass WDs.}
\tablefoottext{b}{Ignition outside the central ball containing all \isotope{56}{Fe}.}
}
\end{table*}

\subsection{Sub Chandrasekhar-mass detonation}\label{s:scfe}

In the following sections, it is analysed the nucleosynthesis resulting from the detonation of bare CO WDs of solar metallicity in which \isotope{56}{Fe} is concentrated in a central ball with different mass fractions. The models reported here adopt different values for the abundance of \isotope{56}{Fe} at the centre of the WD: $X_\mathrm{c}(^{56}\mathrm{Fe})=10^{-3}$, which is roughly the Solar System fraction of \isotope{56}{Fe} and represents the case with no iron sedimentation at all, $X_\mathrm{c}(^{56}\mathrm{Fe})=0.35$, which is representative of the sedimentation of a COFe alloy as described by \citet{2021ApJ...919L..12C}, $X_\mathrm{c}(^{56}\mathrm{Fe})=0.50$ and 0.70, both cases belonging to an intermediate phase in the process of aggregation of a pure iron core. 
In all cases, the total mass of \isotope{56}{Fe} is the same and represents a fraction $X(^{56}\mathrm{Fe})=1.05\times10^{-3}$ of the WD mass. The metallicity is $Z=0.014=X(^{56}\mathrm{Fe})+X(^{22}\mathrm{Ne})$ and the abundance of \isotope{22}{Ne} is homogeneous through the WD.
Table~\ref{tab3} shows the models computed and the outcome of the explosion. As in the case of sedimentation of \isotope{22}{Ne}, the kinetic energy and the yield of \isotope{56}{Ni} are nearly insensitive to the sedimentation of \isotope{56}{Fe}. 

\begin{figure}
\resizebox{\hsize}{!}{\includegraphics{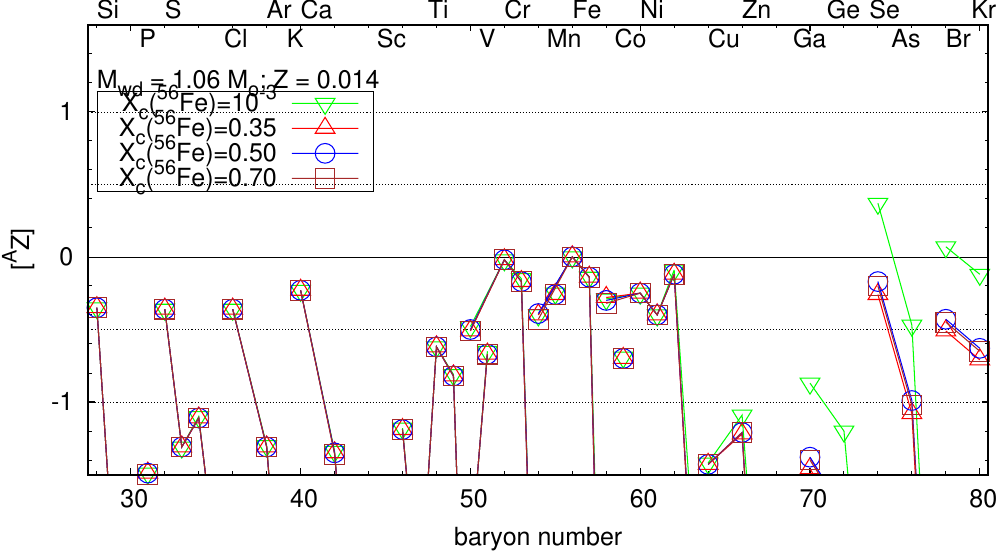}}
\caption{Nucleosynthesis of sub Chandrasekhar-mass bare CO detonation models with $M_\mathrm{WD}=1.06$~M$_\sun$ and $Z=0.014$ with different central concentrations of \isotope{56}{Fe} (see Table~\ref{tab3} for details). The abundance of each isotope is given with reference to the final yield of \isotope{56}{Fe} and normalized to Solar System abundances, as in Fig.~\ref{f:a}.
}
\label{f:h}
\end{figure}

Figure~\ref{f:h} shows the production factors of the sub Chandrasekhar-mass models with different central concentrations of \isotope{56}{Fe}. As it can be seen, the yield of every single isotope up to a baryon number $A=70$ is unaffected by the separation of \isotope{56}{Fe}. Furthermore, the yields of the isotopes with $A>70$ that are affected by such process are independent of the degree of concentration of \isotope{56}{Fe} in the centre as long as it is absent from the WD mantle. The main changes in the nucleosynthesis are the reductions of the yields of selenium, gallium, germanium, krypton, and arsenic, which decrease down to a factor four, although the impact is uneven for different isotopes of each element. Most of these species, in particular the p-nuclei \isotope{74}{Se} and \isotope{78}{Kr}, are synthesized during the explosion when the detonation front arrives to the outermost layers of the WD, approximately 0.03~M$_\sun$ below the surface, burning up to a temperature of $2.8-3.2$~GK at a density $\sim1-3\times10^6$~\gccb \citep[A comprehensive description of the p-process in SNe Ia can be found in, e.g.,][]{2011tra}. They are the result of a $\gamma$-process starting from seed \isotope{56}{Fe} and are most affected by the migration of such isotope to the central layers of the WD. 

\subsection{Chandrasekhar-mass WD delayed detonation}\label{s:cfe}

\begin{figure}
\resizebox{\hsize}{!}{\includegraphics{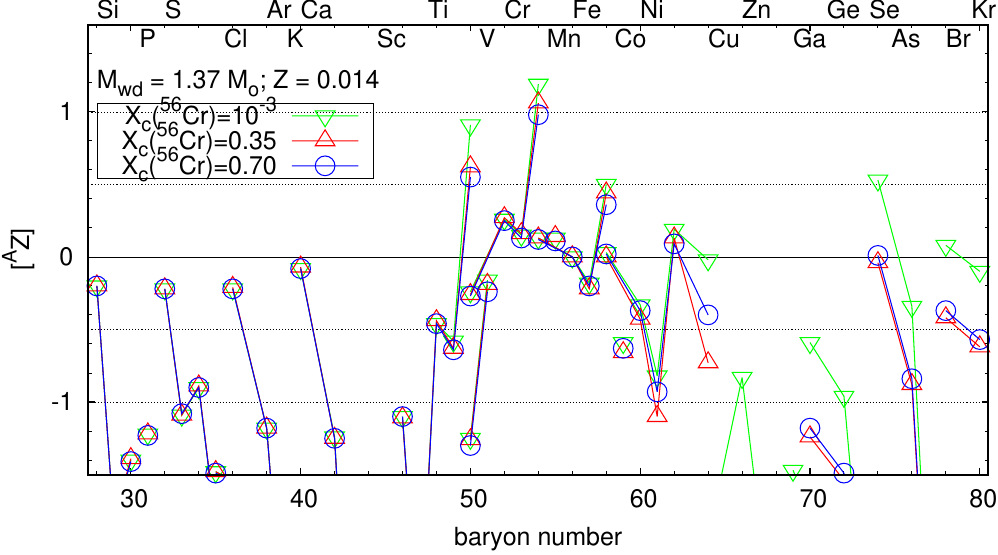}}
\caption{Same as Fig.~\ref{f:h}, but for the Chandrasekhar-mass delayed-detonation models with $\rho_\mathrm{c}=3.5\times10^9$~\gccb and $Z=0.014$.
}
\label{f:f}
\end{figure}

Figure~\ref{f:f} shows the nucleosynthesis of the Chandrasekhar-mass models with a central density $\rho_\mathrm{c}=3.5\times10^9$~\gccb and the different degrees of \isotope{56}{Cr}\footnote{As explained in Sect.~\ref{ss:cf}, all \isotope{56}{Fe} is converted to \isotope{56}{Cr} at the centre of Chandrasekhar-mass WDs because of the high density.} concentration at the centre detailed in Table~\ref{tab3}. Two different trends can be identified in the plot concerning the changes induced by \isotope{56}{Fe} distillation. On the one hand, there are the $A>70$ isotopes, made on the external layers of the WD, whose abundance drops sharply if iron is concentrated on the centre of the WD, as in the sub Chandrasekhar-mass scenario. 

On the other hand, there are isotopes synthesized close to the centre of the massive WD whose abundance is sensitive to the degree of concentration of \isotope{56}{Cr} in the central layers. Some examples are: \isotope{66}{Zn}, \isotope{64}{Ni}, \isotope{54}{Cr} and \isotope{50}{Ti}, whose abundances are reduced by factors 2-7 with respect to the homogeneous model. 
The decrements in the abundances of these isotopes can be traced back to the small amount of nuclear energy released by burning \isotope{56}{Cr}. For instance, a mixture of 50\%-25\%-25\% \isotope{56}{Cr}-\isotope{12}{C}-\isotope{16}{O} releases less than 60\% energy as one made of 50\%-50\% \isotope{12}{C}-\isotope{16}{O}. Less energy released means a lower temperature at NSE and a reduced capacity to create light nuclei, such as protons. Since protons are the main contributors to neutronization by electron capture in NSE \citep{1986thi,2000brc}, the final electron mole number is higher starting from the \isotope{56}{Cr} contaminated mixture than when this species is absent.

It has also been examined if the impact of \isotope{56}{Fe} sedimentation changes significantly with the central density of a massive WD by repeating the same models starting from $\rho_\mathrm{c}=5.0\times10^9$~\gcc, also summarized in Table~\ref{tab3}. The results are almost the same as for $\rho_\mathrm{c}=3.5\times10^9$~\gcc, 
with the only relevant exception of \isotope{48}{Ca} that is strongly overproduced in the homogeneous model at the highest density and whose abundance suffers a decline of up to one order of magnitude if the explosion occurs after all \isotope{56}{Fe} migrates to the centre of the WD.

\section{Explosion of a WD after distillation of \isotope{56}{Fe} and \isotope{22}{Ne}}\label{s:tot}

\begin{table}
\caption{
SNe Ia models after distillation of \isotope{56}{Fe} and \isotope{22}{Ne}.
}\label{tab4}
\centering
\begin{tabular}{rrrrr} 
\hline\hline  
Model & $K$ & $M(^{56}\mathrm{Ni})$ & $M(^{55}\mathrm{Mn})$ & $M(^{58}\mathrm{Ni})$ \\
& ($10^{51}$~ergs) & (M$_{\sun}$) & ($10^{-2}\mathrm{M}_{\sun}$) & ($10^{-2}\mathrm{M}_{\sun}$) \\
\hline
uniform & 1.339 & 0.649 & 0.610 & 4.23 \\
stratified & 1.338 & 0.633 & 0.257 & 6.62 \\
\hline
\end{tabular}
\tablefoot{
All models assumed a WD mass $M_\mathrm{WD}=1.06$~M$_{\sun}$ and progenitor metallicity $Z=0.038$.
}
\end{table}

\begin{figure}
\resizebox{\hsize}{!}{\includegraphics{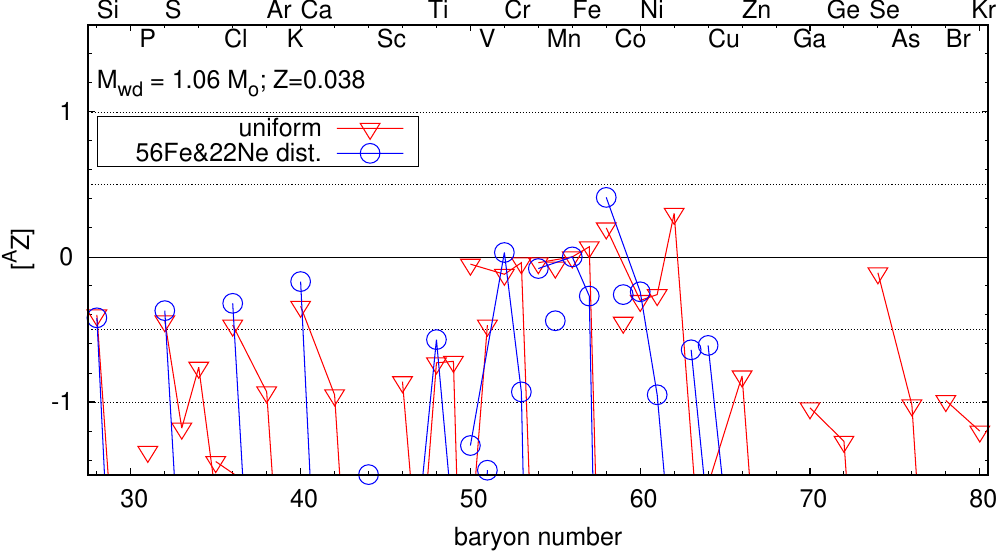}}
\caption{
Nucleosynthesis of sub Chandrasekhar-mass bare CO detonation models with $M_\mathrm{WD}=1.06$~M$_\sun$ and $Z=0.038$, either with uniform chemical composition through the WD or chemically stratified after migration of \isotope{56}{Fe} and \isotope{22}{Ne} to the centre of the WD. The abundance of each isotope is given with reference to the final yield of \isotope{56}{Fe} and normalized to Solar System abundances, as in Fig.~\ref{f:a}.
}
\label{f:g}
\end{figure}

As a final test, there have been explored the properties of the thermonuclear explosion of a stratified WD in which both \isotope{56}{Fe} and \isotope{22}{Ne} have migrated to the centre forming a structure with three chemically differentiated zones. A 1.06~M$_\sun$ WD in hydrostatic equilibrium has been built with a central ball containing all the \isotope{56}{Fe} of the WD\footnote{The total mass of \isotope{56}{Fe} in the WD is 0.0027825~M$_\sun$, that is a fraction $2.625\times10^{-3}$ of the WD mass, while the total mass of \isotope{22}{Ne} is 0.0371~M$_\sun$.}, with a mass fraction $X(^{56}\mathrm{Fe})=0.35$, surrounded by a shell composed of \isotope{12}{C} and the remaining \isotope{22}{Ne} with the azeotropic composition, that is $X(^{22}\mathrm{Ne})=0.30$, and a mantle made up of just \isotope{12}{C} and \isotope{16}{O}. The central density of such WD is $\rho_\mathrm{c}=5.9\times10^7$~\gcc. A chemically homogeneous WD has been exploded as well, with the same mass of each species as in the stratified model, to serve as a reference model
\footnote{The homogeneous WD differs from that in the first row of Table~\ref{tab1} because the last does not include \isotope{56}{Fe} in the initial model.}. This last WD has a central density $\rho_\mathrm{c}=5.4\times10^7$~\gcc. 

The results are shown in Table~\ref{tab4} and Fig.~\ref{f:g}. A comparison with Tables~\ref{tab1} (uniform and full models in first and third rows) and \ref{tab3} (model with $X(^{56}\mathrm{Fe})=0.35$ in second row), and with Figs.~\ref{f:a} and \ref{f:h} makes it clear that the impact of considering both \isotope{56}{Fe} and \isotope{22}{Ne} sedimentation is equivalent to the superposition of the effects of the sedimentation of each one independently. In spite of \isotope{56}{Fe} occupying the centre of the WD and displacing the \isotope{22}{Ne}-rich region slightly outwards, in our models there is no distinguishable interference among the effects of their sedimentation on the nucleosynthesis result of a SNe Ia explosion. 

\section{Conclusions}\label{s:conclusions}

A one-dimensional supernova code has been used to explore the outcomes of thermonuclear explosions of frozen WDs in which some of the secondary (in terms of abundance) species separate upon crystallization and concentrate either on the centre or on an off-centre shell. To this end, a simple chemical profile has been adopted at the beginning of the explosion, following the findings of the most recent studies of the phase diagram of carbon, oxygen and impurities such as \isotope{22}{Ne} and \isotope{56}{Fe}. Complications such as re-mixing after solid melting or the composition of accreted matter have been disregarded in order to discern what may be the maximum effect of chemical stratification on the properties of SNe Ia.
The most relevant results concerning our models are:
\begin{itemize}
 \item{Models show that the chemical separation of \isotope{22}{Ne} or \isotope{56}{Fe} has a quite limited impact on the main properties of SNe Ia, that is kinetic energy and ejected mass of \isotope{56}{Ni}.}
 \item{The sedimentation of \isotope{22}{Ne} with the azeotropic composition, $X(^{22}\mathrm{Ne})=0.30$, can leave an imprint on the supernova yields. In sub Chandrasekhar-mass explosions, the abundances of cobalt and copper increase and those of manganese and chromium decrease. In Chandrasekhar-mass explosions, the concentration of \isotope{22}{Ne} in the centre of the WD harms the synthesis of most iron-group isotopes and, in particular, that of \isotope{55}{Mn}, which is reduced by a factor four.}
 \item{A high concentration of \isotope{22}{Ne} at the centre of a Chandrasekhar-mass WD can affect the path to explosion, that is the central density at thermal runaway, leading to decrements in the production of \isotope{58}{Ni}, \isotope{54}{Fe} and \isotope{55}{Mn} down to factors two, three and four, respectively.}
 \item{The distillation of \isotope{56}{Fe} to the centre of a WD difficults the synthesis of $\gamma$-process elements in the external layers, as is the case for \isotope{74}{Se} and \isotope{78}{Kr}.} 
 \item{In Chandrasekhar-mass explosions, the yields of neutron-rich species such as \isotope{54}{Cr} can be strongly reduced, depending on the degree of concentration of \isotope{56}{Cr} (the by-product of double electron capture starting on \isotope{56}{Fe}) at the centre at the time of explosion. However, the pre-explosive phase of accretion and simmering associated to Chandrasekhar-mass events has not been explored here. Hence, it cannot be discarded that the low electron mole number of \isotope{56}{Cr} induces some kind of dynamical instability prior to the explosion and dilutes its concentration at the centre of the WD.}
 \item{A self-sustained detonation is possible in iron-rich mixtures with carbon and oxygen at a density of $5\times10^7$~\gccb as long as the \isotope{56}{Fe} mass fraction in the mixture is lower than 0.7.}
 \item{Steady conductive flames can propagate through mixtures rich in iron-group elements at a density of $5\times10^9$~\gccb even when the mass fraction of carbon and oxygen is as low as 0.15 (each one), albeit the flame velocity varies sharply with their initial abundance.}
 \item{The sedimentation of \isotope{22}{Ne} with the azeotropic composition has the potential of changing several signatures of progenitor metallicity (the ratio of calcium to sulphur) and of the character of the progenitor, whether a sub Chandrasekhar-mass or a Chandrasekhar-mass WD (stable nickel profile and presence of a central hole in the distribution of radioactive nickel).}
\end{itemize}

The degree of chemical separation depends on the time necessary to allow the WD to cool down the solidification point and an additional time, that can be quite long, to completely solidify. This means that, depending on the parameters of the explosion scenario a variety a chemical structures can be at work. The main parameters defining the properties of the supernova explosion seem not to be afected, but the use of minor species for diagnostic purposes has to be taken with care and to be analysed case by case.

In the case of \isotope{56}{Fe} sedimentation, the time scale is much shorter because the high atomic number of iron, as compared to carbon and oxygen, allows it to form a strongly coupled plasma at a relatively high temperature \citep{2021ApJ...919L..12C}.
If iron and other high atomic number species present in a WD are able to migrate massively to the centre before a SNe Ia explosion, the outer layers of the ejecta will lack any imprint of their initial metal content, except in the region containing the freshly accreted material. In normal-bright SNe Ia models assuming \isotope{56}{Fe} distillation, the abundance of iron and nickel (whose synthesis in the outer layers requires iron seeds) falls abruptly above $\sim20,000$~\kms. Since the presence of iron-group elements is expected to cause UV line blanketing and efficiently suppress the flux in this wavelength region \citep{2000len}, the distillation of iron may affect the UV and be misinterpreted as if the WD progenitor was a low-metallicity star. For the same reason, if the fraction of SNe Ia coming from frozen WDs is high enough, it may affect current attempts to establish a statistical correlation between the UV flux of these events and their host galaxy metallicity \citep{2020pan,2020bro}.

Frozen chemically-segregated WDs are expected to represent at most a fraction of the progenitors of SNe Ia. The delay time between the formation of a WD and a thermonuclear explosion is not constrained, but it can be in the range few tens of Myrs to several Gyrs. 
If the time it takes a WD to start freezing is $1-3$~Gyrs, applying the delay-time distribution of SNe Ia of \cite{2021cas} it can be estimated that up to $<20-40\%$ of all SNe Ia can explode in a state of partial or total \isotope{22}{Ne} stratification and an even higher fraction may occur after \isotope{56}{Fe} sedimentation.
Given the number of SNe Ia currently detected every year, it can be expected that many of these events occur in, at least, partially chemically-segregated WDs. Therefore, it is important to keep in mind how it can affect the appearance of such events.

\begin{acknowledgements}
E.B. and L.P. acknowledge partial support from the Spanish grant PID2021-123110NB-100 funded by MCIN/AEI/10.13039/501100011033/FEDER/UE. J. I. acknowledges funding from MICIN/AEI grant PID2019-108709GBI00 and program Unidad de Excelencia Maria de Maetzu CEX2020-001058-M. L.P. acknowledges partial financial support from the Italian MUR project2022RJLWHN: Understanding R-process \& Kilonovae Aspects (URKA).
\end{acknowledgements}

 \bibliographystyle{aa}
\bibliography{sn23}
\end{document}